\documentclass[preprint,authoryear,12pt,english]{article}

\usepackage{natbib}
\usepackage{url}
\makeatletter
\def\url@leostyle{%
    \def\UrlFont{\sf}}{\def\UrlFont{\small\ttfamily}}
\makeatother
\usepackage{pst-plot}
\usepackage{pstricks-add}

\defcitealias{kant1781guyer}{CPR}
\defcitealias{kant1790pluhar}{CJ}

\usepackage{graphicx}
\usepackage{babel}
\usepackage{units}

\usepackage{hyperref}
\usepackage{authblk}

\date{}

\begin{document}

\title{The Kantian Framework of Complementarity\footnote{\textbf{Notice:} this
    is the preprint version of a work that has been published
    in Studies in History and Philosophy of Modern Physics as: Cuffaro,
    M., ``The Kantian Framework of Complementarity,'' \emph{Studies in
    History and Philosophy of Modern Physics, 41} (2010),
    pp. 309-317. \href{http://dx.doi.org/10.1016/j.shpsb.2010.04.003}
    {http://dx.doi.org/10.1016/j.shpsb.2010.04.003}. Changes
    resulting from the publishing process, such as peer review, editing,
    corrections, structural formatting, and other quality control
    mechanisms are not reflected in this document. Changes have
    been made to this work since it was submitted for publication.
    A summary of the substantive changes between this document and the
    published version follows: sections 1 (The Kantian Framework)
    and 2 (The Genesis of Quantum Mechanics), below, do not
    appear in the published version. Pedagogical and informal
    examples which are used, in this version of the paper, to
    illustrate physical and philosophical concepts (such as the
    discussion of the fish on p. 9, the geologist on p. 10, the
    sine wave and the boats on p. 12) do not appear
    in the published version. Lastly, the published
    version of the paper contains an extended presentation of the
    views of (and Bohr's relationship with) Bohr's mentor, Harald
    H\o ffding, as well as an extended presentation of Kant's views
    on the regulative ideas of reason and his discussion of the
    positive use of the noumenon (which is primarily found in the
    Critique of Judgement and not in the Critique of Pure Reason).
  } \footnote{I am indebted to Lindsay Doucet, Lori Kantymir, Molly
    Kao, Gregory Lavers, Robert Moir, Kathleen Okruhlik, Morgan Tait,
    Martin Vezer, Vladimir Zeman, and especially William Harper and
    Wayne Myrvold for their helpful comments and criticisms of my
    earlier drafts of this paper.
  }
}

\author{Michael E. Cuffaro}
\affil{The University of Western Ontario, Department of Philosophy}

\maketitle

\thispagestyle{empty}

In the literature on the Copenhagen interpretation of quantum
mechanics, not enough attention has been directed to the similarities
between Bohr's views on quantum mechanics and Kant's theoretical
philosophy. Too often, the connection is either ignored, downplayed,
or denied outright. This has, as far as a proper understanding of
Bohr's views is concerned, been detrimental, for it has contributed to
the common misconception of Bohr as either a positivist or a
pragmatist thinker.\footnote{\citet[]{baggot2004}, for example, does
  not mention Kant at all in relation to Bohr. \citet{folse1985}, on
  the other hand, flatly denies any similarities whatsoever that are
  not merely superficial. Baggot and Folse both view Bohr as a
  pragmatist. For examples of positivist construals of Bohr, see:
  \citet{popper1982}, \citet{bunge1955a, bunge1955b}.} In recent
years, however, there has been a growing number of commentators
attentive enough to note the important affinities in the views of
these two thinkers (for instance, \citealt[]{honner1982},
\citealt[]{mackinnon1982}, \citealt[]{shimony1983},
\citealt[]{kaiser1992}, \citealt{chevalley1994},
\citealt{faye2008}). All of these commentators are, I believe,
correct; however the picture they present to us of the connections
between Bohr and Kant is one that is painted in broad strokes. It is
open to the criticism that these affinities are merely superficial.

The contribution that I intend to make in this essay, therefore, is to
provide a closer, structural, analysis of both Bohr's and Kant's views
that makes these connections more explicit. In particular, I will
demonstrate the similarities between Bohr's argument, on the one hand,
that neither the wave nor the particle description of atomic phenomena
pick out an object in the ordinary sense of the word, and Kant's
requirement, on the other hand, that both `mathematical' (having to do
with magnitude) and `dynamical' (having to do with an object's
interaction with other objects) principles must be applicable to
appearances in order for us to determine them as objects of
experience. I will argue that Bohr's `Complementarity interpretation'
of quantum mechanics, which views atomic objects as idealizations, and
which licenses the repeal of the principle of causality for the domain
of atomic physics, is perfectly compatible with, and indeed follows
naturally from a broadly Kantian epistemological framework.

There are exegetical difficulties with respect to both Bohr and
Kant. Their writings are dense and are considered to be obscure by
many. Interpreting Kant has become something of an industry in
philosophy. As for Bohr, J.S. Bell writes of him: ``While imagining
that I understand the position of Einstein ... I have very little
understanding of the position of his principal opponent, Bohr.''
\citeyearpar[p. 155]{bell1981}. Abner Shimony writes: ``I must confess
that after 25 years of attentive\textemdash and even
reverent\textemdash reading of Bohr, I have not found a consistent and
comprehensive framework for the interpretation of quantum mechanics.''
\citeyearpar[p. 109]{shimony1985}. I do not pretend to have succeeded,
where these and other eminent physicists and philosophers have failed,
in resolving all of the problems that go along with giving a
comprehensive and consistent interpretation of Bohr's philosophical
position. Bohr is known to have thought highly of the Pragmatist
philosophy of William James, and Bohr's philosophy represents, in all
likelihood, a combination of Jamesian and Kantian strands (although
even this is likely an oversimplification). In this essay it is the
Kantian aspects of Bohr's views that I will focus on; I do not,
however, believe this is the whole story.\footnote{For more on Bohr
  and William James, see, e.g, \citet[p. 49-51]{folse1985}.}

Understanding the Kantian aspects of Bohr's thought is important
because, although Bohr's and Kant's philosophies do diverge
ultimately, they nevertheless share (as I will argue) a common
epistemological framework. Any interpretation of Bohr should,
therefore, \emph{start} with Kant. Further, comparing Kant and Bohr is
also invaluable for our interpretation of Kant. By asking the question
`how can a Kantian make sense of quantum mechanics?', one gains
valuable insight into the implications of the principles of quantum
mechanics for Kantian philosophy\textemdash in particular, what the
uncertainty relations, if accepted, entail for the applicability of
Kant's principle of cause and effect.

The essay is structured as follows: section 1 is devoted to a
discussion of Kant's characterization of objective cognition. In
section 2, I review the history of quantum theory up to Heisenberg's
development of the Uncertainty Principle. In section 3, I give an
analysis of Bohr's arguments for Complementarity, and discuss the
Kantian aspects of Complementarity in section 4. Finally, in section
5, I deal with some possible objections to my interpretation of
Bohr. For those already familiar with Kantian philosophy and/or the
early history of quantum mechanics, sections 1 and 2 may be skimmed
over (or skipped) without disastrous results, and referred to as
needed for clarification of the exposition that follows in sections 3
and 4.

\section{The Kantian Framework}

For Kant, there are two aspects to experience: on the one hand, there
are \emph{intuitions}. We subsume these intuitions, on the other hand,
under \emph{concepts}. Intuitions are mediated by \emph{sensibility}:
our mind's capacity to be affected by objects
\citepalias[A19/B33]{kant1781guyer}. The effect, on our sensibility, of
some object is the \emph{sensation} of that object, and
\emph{empirical intuition} is that aspect of an intuition that is
associated with this sensation.

An \emph{appearance} is ``The undetermined object of an empirical
intuition'' \citepalias[A20/B34]{kant1781guyer} (for example, consider a
person in a dark room who sees a shape against the far wall, but only
after some scrutiny determines that shape to be a chair. Before
determining it to be a chair, the person is puzzled as to what it is:
we can say that the person views it merely as the appearance of
something indeterminate).\footnote{For a more thorough discussion of
  this point, see \citet[p. 110-111]{harper1984}.} There are two
aspects to an appearance. First, there is its \emph{matter}: what we
sense. Second, there is its \emph{form}. This is how the matter is
\emph{related}, both to itself and to the subject. The two forms of
appearances are: space, for outer appearances, and time, for both
inner and outer appearances. As forms of appearances, they are the
\emph{formal conditions} for appearances; they are a priori (in a
logical, not a temporal, sense), i.e., they are the necessary
relations according to which sensations must be ordered in our mind
\citepalias[A20/B34]{kant1781guyer}.

So much for intuition. \emph{Concepts of the understanding}, now,
correspond to rules for synthesizing the manifold of intuition. For
example, an empirical concept (e.g., a horse) corresponds to a
rule according to which this bushy tail, that long nose, that mane,
and those hoofs can be associated in one representation. When we
synthesize, i.e., combine, some particular manifold of intuition
according to the particular rule for a concept, we say that this
manifold of intuition has been subsumed under the concept. Now a
\emph{pure} concept of the understanding (a `category') is one of a
set of meta-concepts that all empirical concepts necessarily
presuppose. Like the pure forms of intuition, these categories are a
priori.\footnote{For a list,
  \citetalias[\emph{Cf}.][A80/B106]{kant1781guyer}.}

Associated with the categories are formal principles for their
application to possible experience. Among these, Kant distinguishes
between \emph{mathematical} and \emph{dynamical} principles for the
possibility of experience
\citepalias[B198-B294]{kant1781guyer}.\footnote{The mathematical principles
  are the \emph{Axioms of Intuition} and \emph{Anticipations of
    Perception}; the dynamical principles are the \emph{Analogies of
    experience} and the \emph{Postulates of empirical thought as
    such}. As the Postulates do not have a direct bearing on our
  discussion, I will leave them aside here and focus exclusively on
  the Analogies.} The former are \emph{constitutive for
appearances}. They are necessary principles for the possibility of
presenting an appearance to ourselves as \emph{existing}. These say
that in order for anything to appear to us, it must be apprehended as
having, determinately, both an extensive (length, breadth, etc.) and
an intensive magnitude (i.e., a degree). But that something appears to
us as existing is, by itself, not enough to determine this something
as an \emph{object}. To determine this appearance as an object,
we must apply the dynamical principles to it. The dynamical
principles are not constitutive but \emph{regulative}.\footnote{This
  sense of regulative should not be confused with the sense that
  Kant uses with respect to the `ideas of reason'. There the
  distinction is between that which is constitutive or regulative
  with respect to experience as a whole. Here, he uses regulative
  not in the context of experience in general, but in the context of
  particular objects of experience.} They are principles, not for the
apprehension, but for the \emph{connection} of appearances in time;
they presuppose that an appearance has already been apprehended in
accordance with the mathematical principles. These dynamical
principles state, first, that all change presupposes something
permanent; second, that all change must occur according to the law of
cause and effect; third, that all substances that are perceived as
simultaneous are in mutual interaction. To determine an appearance as
an object of a possible experience, therefore, we require that at a
determinate instant in time, it has a determinate position in space
(determined by the mathematical principles) and that there is a law
(subject to the dynamical principles) by which it dynamically
interacts with other objects.

In particular, the principle of causality tells us, according to Kant,
that in order to cognize change in some object, there must be a rule
by which we objectively associate our perceptions of the object through
time; i.e., some objective ordering of our perceptions such
that the state of the object at some moment in time is presented as
being in a determinate relation with the states of the object at
other times. To illustrate: suppose I lean against a fence at the
bank of a river, and watch a log as it is carried downstream
by the current.\footnote{This is a variation on Kant's example of the ship
  (\emph{Cf}. \citetalias[B236-238]{kant1781guyer}).} At time $t_1$, I
watch as it comes into view from around the bend in the river some yards
upstream. I then daydream for a while. Eventually, I notice ($t_2$)
that the log has travelled some distance from the place where I first
spotted it. At $t_3$, I recall to myself the motion of the log down
the river that I half-consciously observed while daydreaming, after
which I continue to watch the log as it disappears into the forest
($t_4$). Later that afternoon, I recall that what aroused me from
my daydream was a sparrow alighting on the log ($t_5$).
If we list these representations in the order in which they are
actually perceived, then this is a \emph{subjective
ordering}: $$t_1, t_2, t_3, t_4, t_5$$
I can also give these perceptions an \emph{objective ordering},
however, according to which the motion of the log must have actually
proceeded in time: $$t_1,t_3,t_5,t_2,t_4$$ To determine this objective
ordering, I must take into account the position of the log in the
river during each of my perceptions, as well as anything else that is
relevant to the motion of the log. The particular rule of succession
for the change of state of the log is something that can only be
discovered empirically (e.g., by determining which way the river is
flowing). However, \emph{that there is} some rule to be discovered is
what the principle of causality tells us. This is \emph{a
  priori}, according to Kant.

Now, for Kant, the presentations of time and space are continuous,
infinitely divisible, quantities \citepalias[B211]{kant1781guyer}. No
matter how small, every `piece' of space or time always presupposes a
possible further intuition of space or time within its boundaries. The
principle of cause and effect, now, tells us that every series of
perceptions has some objective ordering according to which it
progresses in time. But since time is infinitely divisible, so is the
progression of perceptions \citepalias[B255]{kant1781guyer}. All
\emph{change} associated with a possible experience is continuous,
therefore, and we can know this a priori, according to Kant.

Kant's views were highly influential among both scientists and
philosophers, especially in the latter half of the 19$^{th}$
century. In the 20$^{th}$ century, however, the development of Special
and General Relativity, and the development of Quantum Mechanics
undermined Kant's views in the eyes of many. Relativity theory called
into question the a priori status of space and time; Quantum Mechanics
called into question the a priori status of both space and time and
the principle of causality. It is to the latter theory that we now
turn.

\section{The Genesis of Quantum Mechanics}

When certain substances are subjected to extremely high temperatures,
they absorb energy and emit light (picture a blacksmith hammering
metal in a forge; if the temperature is high enough, the metal becomes
`red hot' or `white hot'). To study this phenomenon, 19$^{th}$ century
physicists conceived of the notion of a `black body': a completely
non-reflecting object capable of emitting radiation at an intensity
directly proportional to the amount of energy it absorbs. In 1859-60,
Kirchoff was able to show that the ratio of emitted to absorbed energy
in these materials depended solely on frequency and temperature
(and not on, e.g., the body's shape).

At the turn of the century, Planck hypothesized that the total energy
of the system was distributed over a large collection of
indistinguishable energy elements. The result was the now famous
formula: $\varepsilon = h\nu$. That is, the energy, $\varepsilon$, in
each element is equal to the constant of proportionality, $h$
(Planck's constant), times the frequency, $\nu$. His results implied
that $\varepsilon$ must be given in \emph{integer multiples} of
$h\nu$. This was unprecedented (in classical physics, physical
quantities change continuously with time). Heisenberg speculates:

\begin{quote}
... he must soon have found that his formula looked as if the
oscillator could only contain discrete quanta of energy\textemdash a
result that was so different from anything known in classical
physics that he certainly must have refused to believe it in the
beginning. ... Planck must have realized at this time that his
formula had touched the foundations of our description of nature
... \citep[p. 35]{heisenberg1959}.
\end{quote}

The discovery of Planck's constant (the `quantum of action') was key
to the resolution of some other outstanding problems in physics at the
time. In 1905, Einstein described light in terms of energy
quanta. He later did the same for atoms and ions. In 1913, Bohr's
theory of the atom was published, according to which electrons are
confined to fixed orbits that depend on `principal quantum
numbers'. In his PhD thesis of 1924, de Broglie demonstrated the
wave-particle duality of matter by relating Planck's relation
$\varepsilon = h\nu$ with Einstein's mass-energy equivalence relation,
$\varepsilon = mc^2$.

Schr\"odinger developed his famous wave function for the evolution of
a quantum mechanical system in 1925. Its interpretation was the
subject of some debate. On Schr\"odinger's view, the wave function
represented a real disturbance in the electromagnetic field;
elementary `particles' were really just electromagnetic waves of
different amplitudes, phases, and frequencies, which when
superimposed, resulted in large, localized, waves (called `packets')
which gave the impression of particles moving through space and
time. As Lorentz pointed out, however, wave packets persist only when
they are large compared with their wavelength. When confined to small
regions of space, wave packets, unlike elementary particles, disperse
rapidly \citep[\emph{Cf}.][pp. 214-217]{moore1992}.

Heisenberg wrote his famous uncertainty paper in 1927, in which he
argued for the in principle impossibility of precisely determining
\emph{both} the position and momentum of an elementary particle at any
one time. Presupposing a particle interpretation of elementary
objects, he presented the following thought experiment: imagine we
wish to determine the position and momentum of an electron as it
travels under a microscope. To determine position, we use
high-frequency $\gamma$-rays, since the resolving power of the
microscope is directly proportional to the frequency of the
beam. Frequency, however, is directly proportional to energy. Now when
a high energy photon (a light particle) collides with an electron, the
electron is knocked off its path (the `Compton effect'). But this
means that making an accurate determination of the electron's position
renders us incapable of accurately determining its momentum: to
determine its momentum we require the position of the electron at two
points along its path, but since the path has been altered by the
first position determination, we cannot determine where the electron
would have been had we not interfered with it.

We can avoid the Compton effect by using low frequency
$\gamma$-rays. But recall that the resolving power of the microscope
is directly proportional to the frequency of the beam. If we use low
frequency photons to measure the \emph{momentum} of the electron, then
we lose the ability to measure its \emph{position}
accurately. Heisenberg showed, mathematically, that an \emph{exact}
determination of the position of the electron resulted in an
\emph{infinite} uncertainty in its velocity (and hence momentum), and
vice versa (he also demonstrated an analogous uncertainty relationship
between energy and time). In other words, the more certain we are of
one parameter, the less certain we are of the other. In the limit,
i.e., as the uncertainty in the determination of one parameter
approaches 0, the uncertainty in the determination of the other
parameter approaches infinity.

\section{Complementarity}

Bohr accepted the validity of the uncertainty relations, but disagreed
with Heisenberg over their significance. For (the young) Heisenberg,
the uncertainty relations represent an epistemic limitation on what we
can know of some object; we presuppose, however, that in spite of this
limitation, the object is perfectly determinate in itself\textemdash
that it is a particle, in fact. For Bohr the significance of the
uncertainty relations is deeper; it is not epistemic in this sense but
rather \emph{conceptual}. For Bohr the uncertainty relations express
the fact that the fundamental `classical concepts' which both the
particle and wave description of elementary objects presuppose
(spatiotemporal concepts, on the one hand, and dynamical concepts, on
the other) are inapplicable in the atomic domain, and that therefore a
definition of the object in terms of these parameters is
precluded. Let us work our way towards this conclusion. In his Como
paper, Bohr writes:

\begin{quote}
... [quantum theory's] essence may be expressed in the so-called
quantum postulate, which attributes to any atomic process an essential
discontinuity, or rather individuality, completely foreign to the
classical theories and symbolised by Planck's quantum of
action. \citeyearpar[p. 580]{bohr1928}.
\end{quote}

Contrasted with the classical theories, here, is the irreducibly
`discrete' nature of atomic processes; the fact that, according to
quantum theory, the observed state of an elementary object
changes discontinuously with time. What this implies, Bohr goes on to
say, is that in our observations of the results of experiments, the
interaction with the `agency of observation' (i.e., the experimental
apparatus) is an ineliminable part of our description of phenomena
(in his later writings, Bohr calls this an ``essential wholeness.''
\emph{Cf.} \citeyear[p. 72]{bohr1954}).

That last step may seem like an inferential leap, but it is
comprehensible in light of our earlier discussion of Kant. We, saw,
with Kant, how the infinite divisibility of time implies that all
change must be continuous. Bohr's argument tacitly makes use of this
assumption. Thus, on the classical conception of nature, change is
continuous. Yet the state transitions of elementary objects are
irreducibly discontinuous. It follows from this that, from a classical
point of view, something is `missing' from our description. What is
`missing', according to Bohr, is a clean distinction between the
experimental apparatus and the object of our investigations; the
`agency of observation' is, in some sense, \emph{a part} of what we
observe.

Is it the case, then, that quantum mechanical descriptions of
phenomena are not objective? No, quantum mechanical descriptions of
phenomena, like classical descriptions, are objective. However what is
different is that for the classical (but not for the quantum) case it
is always possible to determinately describe (and correct for) the
\emph{interaction} between apparatus and object. Suppose I wish to
describe a fish swimming in the water below my motorboat. There are
three components involved in my description: first, there is the
apparatus (my eyes); second, there is the object (the fish); third,
there is the interaction between my eyes and the fish. We describe
this last component by means of light rays that reflect off the fish
and travel through water, then air, and finally into my eyes. Now when
I look at the fish, it appears displaced from its actual position in
the water due to the refraction of the ray. However, in my description
of the fish, I am able to describe the interaction between my eyes and
the fish and I am able to compensate for this interaction in my
description; I am, at least vaguely, aware of the laws for the
refraction of light, and taking these into account, I am able to
determine the actual position of the fish, as well as its movements,
with reasonable certainty; I am able to distinguish the fish `as it
really is' (the object) from the fish as it appears (the
phenomenon). But this is \emph{not} possible for atomic
phenomena. Although we must make some `subject-object'
distinction\textemdash some `cut' in what we observe\textemdash it is
an \emph{arbitrary} cut\textemdash one in which the interaction
between apparatus and object cannot be disentangled from our
description of the object.

One might object that there is some arbitrariness to the cut we make
in the classical realm as well; a geologist and an archaeologist, for
instance, will have distinct objects of inquiry even though both
observe the same physical stone. What is different is that in the
classical case, as we correct for the interaction with the apparatus
in our description of the object, we are constrained by the (according
to Bohr) criteria for its independent reality: a precise location in
space-time and a precise account of its interaction with other
objects. In the quantum case, however, we cannot account for the
interaction with the apparatus in a way that leaves the object with
definite position/time and momentum/energy parameters. Our language
nevertheless requires some distinction, so we arbitrarily impose one.

Bohr expresses all of the foregoing in the following concise
paragraph:

\begin{quote}
  Now the quantum postulate implies that any observation of atomic
  phenomena will involve an interaction with the agency of observation
  not to be neglected. Accordingly, an independent reality in the
  ordinary physical sense can neither be ascribed to the phenomena nor
  to the agencies of observation. After all, the concept of observation
  is in so far arbitrary as it depends upon which objects are included
  in the system to be observed. Ultimately every observation can of
  course be reduced to our sense perceptions. The circumstance,
  however, that in interpreting observations use has always to be made
  of theoretical notions, entails that for every particular case it is
  a question of convenience at what point the concept of observation
  involving the quantum postulate with its inherent `irrationality' is
  brought in. This situation has far-reaching consequences. On one
  hand, the definition of the state of a physical system, as
  ordinarily understood, claims the elimination of all external
  disturbances. But in that case, according to the quantum postulate,
  any observation will be impossible, and above all, the concepts of
  space and time lose their immediate sense. On the other hand, if in
  order to make observation possible we permit certain interactions
  with suitable agencies of measurement, not belonging to the system,
  an unambiguous definition of the state of the system is naturally no
  longer possible, ... \citeyearpar[p. 580]{bohr1928}.
\end{quote}

Bohr must still explain exactly why the classical concepts are not
applicable to elementary objects. He writes:

\begin{quote}
The fundamental contrast between the quantum of action and the
classical concepts is immediately apparent from the simple formulae
which form the common foundation of the theory of light quanta and of
the wave theory of material particles. If Planck's constant be denoted
by $h$, as is well known, $$E\tau = I\lambda = h,
\quad.\quad.\quad.\quad(1)$$ where $E$ and $I$ are energy and
momentum respectively, $\tau$ and $\lambda$ the corresponding period
of vibration and wave-length. In these formulae the two notions of
light and also of matter enter in sharp contrast. While energy and
momentum are associated with the concept of particles, and hence may
be characterised according to the classical point of view by
definite space-time co-ordinates, the period of vibration and
wave-length refer to a plane harmonic wave train of unlimited
extent in space and time. \citeyearpar[p. 581]{bohr1928}.
\end{quote}

In other words, in each case (i.e., for light and matter), Planck's
constant relates two incompatible quantities. In the first relation,
$E$ (energy) is associated with the concept of a particle given with
definite spatiotemporal coordinates, while $\tau$ (period of
vibration) is associated with a wave-train `of unlimited extent', not
conceptualizable with respect to definite space-time coordinates. The
case is the same for $I$ and $\lambda$.

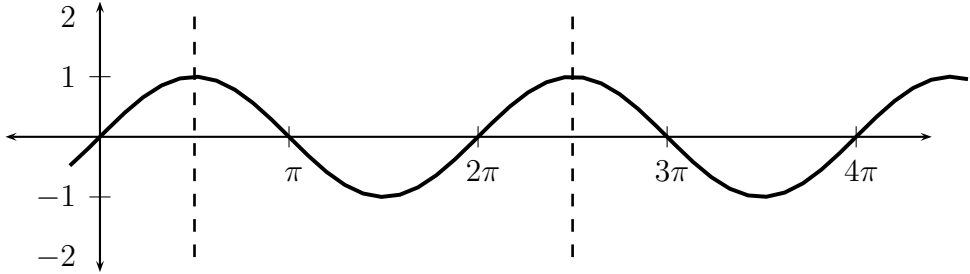
\psset{unit=0.8cm}
\begin{pspicture}(-0.5,-3.25)(15,3.25)
  \psplot[linewidth=1.5pt]%
         {-0.5}{14.424777961}{x 180 mul 3.141592654 div sin}
  \psaxes[xunit=3.141592654,showorigin=false,trigLabels]{<->}(0,0)(-0.5,-2.25)(4.4,2.25)
  \psset{linestyle=dashed,linewidth=1pt}
  \psline[dash=5pt 5pt](1.5707963266,-2)(1.5707963266,2)
  \psline[dash=5pt 5pt](7.8539816330,-2)(7.8539816330,2)
\end{pspicture}

To illustrate the concept of a `wave-train', consider the sine
function. An individual wave (e.g., the one delimited by the two
dashed lines in the figure) is a section of this function that
stretches from one crest to the one immediately following it. The
wave-train is made up of all the individual waves, extending
infinitely in both directions along the $x$-axis. Clearly, it is not
located \emph{at} a particular point in space. Bohr's point is that it
does not make sense to picture an object to ourselves that is, as the
above relations express, \emph{both} given at some definite
spatiotemporal location \emph{and} of unlimited extent in space and
time. Nevertheless, physical theory does provide us with the resources
we need to get around this difficulty, whether we assume a wave or
a particle description of the object. The problem, as we shall see, is
that neither description is precise.

For the  case of the wave description, we can do this by using the
superposition principle. Bohr writes:

\begin{quote}
Only with the aid of the superposition principle does it become
possible to obtain a connexion with the ordinary mode of
description. Indeed, a limitation of the extent of the wave-fields in
space and time can always be regarded as resulting from the
interference of a group of elementary harmonic
waves. \citeyearpar[p. 581]{bohr1928}.
\end{quote}

A boat sailing over a smooth lake creates a wave disturbance behind
it. A second boat, travelling alongside, also creates a
disturbance. When the waves meet they intersect and constructively
interfere. The result is one large, combined, wave. This is called
superposition, and the combined wave is called a wave group. When
enough waves are superimposed in just the right way, the resultant
wave group can be very localized, spatiotemporally; if it is so
localized, then we call the group a wave \emph{packet}, and we
represent the velocity of the wave packet by its group velocity. This
is what is behind Schr\"odinger's picture of a wave, manifesting
particle-like properties, moving through space and time. However,
although the superposition principle enables us to construct a
description of an object in this way, it necessarily involves an
element of indeterminacy with regard to that object.

\begin{quote}
Rigorously speaking, a limited wave-field can only be obtained by the
superposition of a manifold of elementary waves corresponding to all
the values of $\nu$ and $\sigma_x$, $\sigma_y$, $\sigma_z$. But the
order of magnitude of the mean difference between these values for two
elementary waves in the group is given in the most favourable case by
the condition $$\Delta t \Delta \nu = \Delta x \Delta \sigma_x =
\Delta y \Delta \sigma_y = \Delta z \Delta \sigma_z = 1\qquad\qquad
[1a]$$ where $\Delta t, \Delta x, \Delta y, \Delta z$ denote the
extension of the wave-field in time and in the direction of space
corresponding to the co-ordinate axes. \citep[p. 581]{bohr1928}.
\end{quote}

Here, $\nu$ refers to the frequency, and $\sigma_x, \sigma_y,
\sigma_z$ refer to the wavenumbers for the elementary waves in the
directions of the coordinate axes. Exactly how the waves
constructively (or destructively) interfere depends, in part, on the
wavenumbers/frequencies associated with the individual waves in the
wave group. All else equal, the broader the range of wavenumbers in
the group, the more spatially localized the resultant packet will be,
and vice versa. This is what the expression (1a) is telling us; i.e.,

\begin{eqnarray*}
  & \Delta x \Delta \sigma_x = 1 \\
  \mbox{\emph{implies}:} \quad & \Delta \sigma_x = \frac{1}{\Delta x}
\end{eqnarray*}

It is the same for frequency and time. Now, according to the de
Broglie relations, $E = \hbar\nu$, $I = \hbar\sigma$, where $E$ and
$I$ are energy and momentum respectively, and $\hbar = \frac{h}{2\pi}$
is the reduced Planck's constant. If we multiply equation (1a) by
$\hbar$, this gives us the uncertainty relations: $$\Delta t 
\Delta E = \Delta x \Delta I_x = \Delta y \Delta I_y = \Delta z \Delta
I_z = \hbar\qquad\qquad(2)$$ which give the upper bound on the
accuracy of momentum/position determinations with respect to the
wave-field.

Thus, as the wave-field associated with the object gets smaller
\textemdash as we `zoom in', so to speak, on its position and time
coordinates\textemdash the possibility of precisely defining the
energy and momentum associated with the object decreases in
proportion. And the opposite is also true: in order to determine the
object's momentum (or energy), we require a larger
wave-field\textemdash we need to `zoom out'\textemdash but this
foregoes a precise determination of the object's position. `Zooming
in' and `zooming out', however, are associated with different
experimental arrangements. For the case of the $\gamma$-ray
microscope, they are associated with the finite size of the
microscope's aperture; the uncertainty in the position and momentum of
the electron arises, not because of the interaction between two
determinate entities (a photon and an electron), but rather because
certain experimental arrangements, well-suited for precisely
determining momentum, preclude \emph{the definition} of the object in
terms of continuously changing spatiotemporal coordinates, and vice
versa.

\begin{quote}
Indeed, a discontinuous change of energy and momentum during
observation could not prevent us from ascribing accurate values to the
space-time co-ordinates, as well as to the momentum-energy components
before and after the process. The reciprocal uncertainty which always
affects the values of these quantities is, as will be clear from the
preceding analysis, essentially an outcome of the limited accuracy
with which changes in energy and momentum can be defined, when the
wave-fields used for the determination of the space-time co-ordinates
of the particle are sufficiently small. \citep[p. 583]{bohr1928}.
\end{quote}

Thus no one experimental setup allows for an \emph{exact} definition
of the object in terms of both quantities. One experiment can, at
most, give us a picture of ``unsharply defined individuals within
finite space-time regions.'' \citep[p. 582]{bohr1928}.

Now let us see if we will have better luck if we begin, instead, with
a particle description of the elementary object. Here, let us consider
two different experimental arrangements, both variations of a
`one-slit' experiment, where we direct a photon at a thin diaphragm (a
metal plate) into which an opening, or `slit', has been made. On the
other side of the diaphragm is a photographic plate which registers
the light pattern that results. In one version of the experiment,
designed to detect the particle's momentum, the diaphragm is not
rigidly attached to the experimental apparatus, i.e., upon collision
with the particle, the diaphragm will recoil slightly. When we direct
the photon at the diaphragm, as it passes through the slit it will
exchange momentum with the apparatus, which we can measure by the
amount of recoil we observe in the diaphragm. However the recoil of
the diaphragm gives rise to a corresponding uncertainty with regard to
the position of the particle as it passes through the slit (the recoil
of the diaphragm makes it impossible to precisely determine the
location of the slit, and hence the particle, at the moment of
impact):

\begin{quote}
... we lose, on account of the uncontrollable displacement of the
diaphragm during each collision process with the test bodies, the
knowledge of its position when the particle passed through the
slit. \citep[p. 698]{bohr1935}.
\end{quote}

On the other hand, suppose the diaphragm is rigidly fixed to the rest
of the apparatus. In this case, as the photon passes through the slit,
whatever momentum it exchanges with the diaphragm is completely
absorbed by the apparatus\textemdash we thus lose the ability to make
use of this momentum value in order to predict the location of the
particle's impact on the photographic plate. Like the case for the
wave picture, then, we have on our hands two experimental
arrangements, one of which is compatible with a precise
position determination; the other compatible with a precise
momentum determination; however each of these \emph{excludes} the
other.

\section{A Kantian View of Complementarity}

Let us stop and reflect; consider the result of some experiment, say
the mark on a photographic plate. The mark itself is a classical
object. It has definite spatiotemporal coordinates, and it causally
interacts in a definite way with its surroundings. However,
\emph{this} description of the phenomenon\textemdash of the mark
\emph{as a mark} on a photographic plate and nothing more\textemdash
includes the photographic plate. To go further and describe the mark
as a mark that has been left by some \emph{independently existing
object} that has interacted with the plate is what we desire to do,
for this allows us to unify the marks resulting from different
experiments as being different manifestations of the same
independently existing object. Our goal is to `get at'
reality\textemdash the thing behind the phenomena\textemdash as it
exists independently of the conditions of our experiments. We do this
by eliminating the interaction between apparatus and object from our
description of the latter.

Now from a Kantian perspective, in order to describe the object behind
the phenomena as some independently existing\footnote{I.e., in the
  sense of its being the same object in different experimental
  contexts. We can never abstract, on Kant's view, \emph{completely}
  from the subjective conditions of observation (space and time), of
  course.} object of possible experience, we must ascribe, to the
object, first: a determinate position, constrained by Kant's
mathematical principles: the object must have a \emph{definite}
spatial extent and degree; second, a determinate momentum or `quantity
of motion', constrained by Kant's dynamical principles, telling us how
the object interacts with its surroundings\textemdash in particular,
how it \emph{changes} through time.

To visualise the object, we make use, say, of the superposition
principle. But by this means it is impossible to obtain \emph{both} an
exact position and an exact momentum determination (likewise for
energy and time). It is possible to obtain an exact position
determination, but in that case we completely forego a determination
of the particle's momentum, and vice versa. We can, however, get
something like a `complete' object (i.e., one in which both causal
and spatiotemporal parameters are present) by making our position and
momentum determinations \emph{inexact}\textemdash ``unsharply
defined''. But in that case, although our description is objective, it
is no longer the description of an object of possible experience
(i.e., something physically real), for Kant\textemdash for in order
for it to be physically real, we must assign determinate values to
both parameters. Instead, the object is what Kant calls a noumenon, or
abstract object.

To clarify: according to Kant, a concept of the understanding must be
understood both in terms of its form and in terms of the content to
which it can be applied. We can think of the form of a concept as
analogous to a mathematical function, e.g., $f(x) = 2x + 4$. Now a
determinate result can be obtained for this function only if something
is filled in for $x$. By itself, the function only represents a form
for the determination of a variable. Likewise for a concept: without
\emph{determinate} content, a concept gives us no \emph{determinate}
cognition. ``Without [an object] it has no sense, and is
entirely empty of content, even though it may still contain the logical
function for making a concept out of whatever sort of \emph{data}
there are.'' \citepalias[B298]{kant1781guyer}.

The concept of a noumenon is the concept of something
\emph{indeterminate}\textemdash analogous to $x$ in the mathematical
equation. The function above \emph{cannot} be applied to $x$ itself,
but only to a value that has been filled in for $x$. Similarly for
concepts: cognition of an object of possible experience requires that
a concept be applied to a determinate, not indeterminate,
intuition.
A concept of some causal
mechanism corresponds to a rule for the progression of perceptions
\emph{in time}, and the concepts of the understanding, in general,
correspond to rules that must be applied to \emph{our} sensible forms
of intuition, space and time, which are always given
determinately.

But now consider an elementary particle. According to the uncertainty
relations, it is impossible \emph{in principle} to describe the
particle's momentum with any degree of precision without a
corresponding loss of precision with regards to its spatiotemporal
coordinates. It follows that in order to describe it using \emph{both}
position/time (spatiotemporal) and momentum/energy (dynamical)
parameters, the \emph{spatiotemporal} parameters associated with it
must be made \emph{indeterminate}. In fact, both the spatiotemporal
and dynamical parameters must be made indeterminate, but it is the
fact that the spatiotemporal parameters must be made indeterminate
that is the key, for now, on a Kantian picture, the dynamical
principles (whether or not we ascribe determinate dynamical
parameters) are strictly speaking no longer applicable, for the
dynamical principles always presuppose a \emph{determinate} appearance
in space and time apprehended in accordance with the mathematical
principles. The upshot of all of this is that since there is no
determinate spatiotemporal magnitude to apply the dynamical principles
to, we cannot complete our description of the object according to the
Kantian criteria for objects of possible experience. Therefore the
`object' corresponding to our description, on Kant's view, is not
\emph{physically} real.

Bohr reaches the same conclusion regarding the physical reality
of our descriptions of elementary objects:

\begin{quote}
... a sentence like ``we cannot know both the momentum and the
position of an atomic object'' raises at once questions as to the
physical reality of two such attributes of the object, which can be
answered only by referring to the conditions for the unambiguous use
of space-time concepts, on the one hand, and dynamical conservation
laws, on the other hand. \citeyearpar[p. 211]{bohr1949}.
\end{quote}

The issue is not the existence of atomic objects as such (it is undeniable that something gives rise to the phenomena we observe), but
whether our fundamental spatiotemporal and dynamical concepts are
literally applicable to them. Evidently, according to both Bohr and Kant, they are not. And yet these `ordinary' concepts, for
Bohr, are also \emph{necessary} concepts. The experimental apparatus
(a voltmeter, say) is always a piece of classical equipment which
communicates classical information about what we assume to be (using
classical criteria) an independently existing object. The concept of
observation itself, therefore, presupposes the classical concepts.

\begin{quote}
Here, it must above all be recognized that, however far quantum
effects transcend the scope of classical physical analysis, the
account of the experimental arrangement and the record of the
observations must always be expressed in common language supplemented
with the terminology of classical physics. \citep[p. 313]{bohr1948}.

\vspace{12pt}

The main point here is the distinction between the \emph{objects}
under investigation and the \emph{measuring instruments} which serve
to define, in classical terms, the conditions under which the
phenomena appear. \citep[pp. 221-222]{bohr1949}.
\end{quote}

We require the classical concepts, not only to observe, but also to
communicate experimental results:

\begin{quote}
... the requirement of communicability of the circumstances and
results of experiments implies that we can speak of well defined
experiences only within the framework of ordinary concepts.
\citep[p. 293]{bohr1937}.
\end{quote}

The situation seems hopeless. We require the classical criteria in
order to observe a physical object and to communicate the experience;
yet, the classical criteria cannot fulfil their intended function in
the atomic domain, for they mutually exclude each other. Ironically,
it is the uncertainty relations that save us. They guarantee that we
\emph{can} nevertheless achieve a unified description by `patching
together' the mutually exclusive dynamical and spatiotemporal
descriptions of the object under different experimental
conditions. ``The apparently incompatible sorts of information about
the behaviour of the object under examination which we get by
different experimental arrangements can clearly not be brought into
connection with each other in the usual way, but may, as equally
essential for an exhaustive account of all experience, be regarded as
``complementary'' to each other.'' \citep[p. 291]{bohr1937}. The
uncertainty relations guarantee that a causal description can never
contradict a spatiotemporal description\textemdash that the two can be
used in a complementary way\textemdash for any experiment intended to
\emph{determinately} establish the object's spatiotemporal coordinates
\emph{can tell us nothing} about its dynamical parameters, and vice
versa.

\begin{quote}
the proper r\^ole of the indeterminacy relations consists in assuring
quantitatively the logical compatibility of apparently contradictory
laws which appear when we use two different experimental arrangements,
of which only one permits an unambiguous use of the concept of
position, while only the other permits the application of the concept
of momentum ... \citep[p. 293]{bohr1937}.
\end{quote}

We are not licensed, however, to take the next step and ascribe
physical reality to this `patched together' object of our
descriptions, for the object is not real but abstract, and its
classical attributes are idealizations.

\begin{quote}
From the above considerations it should be clear that the whole
situation in atomic physics deprives of all meaning such inherent
attributes as the idealizations of classical physics would ascribe to
the object. \citep[p. 293]{bohr1937}.
\end{quote}

It is not too difficult to make sense of this from a Kantian point of
view. Again, the concept of a noumenon is the key\textemdash this time
in its positive signification as an idea, or concept of reason. Kant
distinguishes two kinds of concepts: ``Concepts of reason serve for
\textbf{comprehension}, just as concepts of the understanding serve
for \textbf{understanding} (of perceptions).''
\citepalias[A311/B367]{kant1781guyer}.

Concepts of reason, or ideas, have no validity with respect to the
cognition of an object of possible experience\textemdash the cognition
of such objects must always refer to determinate (sensible) conditions
according to which they can be given to us in
experience. Nevertheless, these concepts can be used regulatively, to
\emph{connect} the understanding's concepts\textemdash in our case,
the various descriptions of phenomena (the `marks') observed in the context of
individual experiments\textemdash together in a coherent way in the
context of our overall experience.

\begin{quote}
All other pure concepts the critique relegates to the ideas, which are
transcendent for our theoretical cognitive power, though that
certainly does not make them useless or dispensable, since they serve
as regulative principles: they serve, in part, to restrain the
understanding's arrogant claims, namely, that (since it can state a
priori the conditions for the possibility of all things it can
cognize) it has thereby circumscribed the area within which all things
in general are possible; in part, they serve to guide the
understanding, in its contemplation of nature, by a principle of
completeness\textemdash though the understanding cannot attain this
completeness\textemdash and so further the final aim of all
cognition. \citepalias[p. 167-168]{kant1790pluhar}.
\end{quote}

The classical concepts, when they transcend possible experience,
become ideas\textemdash they become the classical idealizations at the
heart of the mechanistic conception of nature. (\emph{Cf.}
\citetalias[\textsection \textsection 69-78]{kant1790pluhar}). In the realm
of atomic physics, however, these dynamical and spatiotemporal
idealizations are incompatible; we cannot use them to describe a
classical object. The uncertainty relations tell us that a precise
determination of one type of parameter entirely excludes any
determination whatsoever of the other type; therefore, they cannot be
used to determine an object of possible experience, which requires a
determination of both. But precisely because they say nothing about the objects of possible experience in this sense, they are compatible with the objects of possible experience\textemdash the
results of our experiments\textemdash just so long as we understand
that when we use these ideas in our description of nature it is only a
manner of speaking; we may only speak `as if' these ideas apply to our
observations.

\begin{quote}
We must be clear that, when it comes to atoms, language can be used
only as poetry. The poet, too, is not nearly so concerned with
describing facts as with creating images and establishing mental
connections (Bohr, quoted in: \citealt[p. 41]{heisenberg1971}).
\end{quote}

Those familiar with Kant should immediately recognize the strategy
being employed here. When confronted with other areas (biology and
ethics, for instance) of human inquiry where the mechanistic
conception (on his view) is either inadequate or
inappropriate, Kant appeals to his doctrine of the antinomies to show
that competing conceptions (e.g., freedom and determinism, mechanism
and teleology) are merely ideas, and that they are compatible with
each other if treated as such (\emph{Cf.} \citetalias[B566-567,
  B586]{kant1781guyer}, \citetalias[\textsection \textsection
  69-78]{kant1790pluhar}).

\section{Concerns and Objections}

One might object that a Kantian should not feel herself committed to
anything like Complementarity, for one may opt to view the
uncertainty relations as an expression of the temporary state of our
ignorance with regard to elementary particles, and not as a final
word. This is correct. A Kantian need not follow Bohr. However, if, as a Kantian, one does accept the uncertainty relations, then something like Complementarity must be the result\textemdash this is what it was my intention to show in this paper. Indeed,
as I have shown, it is because one starts from within the Kantian
framework that the motivation for Complementarity arises. It is
unclear what Kant himself would have thought, but the following
discussion of the mechanistic versus the teleological conceptions of
nature may give us a clue.

\begin{quote}
... I \emph{ought} always to \emph{reflect} on these events and forms
\emph{in terms of the principle} of the mere mechanism of nature,
and hence ought to investigate this principle as far as I can, because
unless we presuppose it in our investigation [of nature] we can have
no cognition of nature at all in the proper sense of the term. But
none of this goes against the second maxim\textemdash that on certain
occasions, in dealing with certain natural forms (and, on their
prompting, even with all of nature), we should probe these and reflect
on them in terms of a principle that differs entirely from an
explanation in terms of the mechanism of nature
... \citepalias[p. 387-388]{kant1790pluhar}.
\end{quote}

Although both the mechanistic and the teleological conceptions are
thought of as `complementary' ideas which guide our investigation of
nature, priority is clearly given, nevertheless, to the mechanistic
conception. The use of the teleological conception is reserved only
for `certain occasions' in which the mechanistic conception is either
inapplicable (perhaps only temporarily) or inappropriate. It is
likely that Kant would have been more conservative than Bohr, i.e.,
that he would not have accepted the uncertainty relations as final. In
that case, one way to interpret Bohr's Complementarity doctrine is as
an attempted refutation of what he took to be Kantian philosophy, with
its overemphasis on the mechanistic conception of nature. Indeed, this
is one way to reconcile Bohr's oft-cited criticisms of `a priorism'
(\emph{Cf.} \citealt[pp. 217-221]{folse1985}) with his insistence on
the bedrock-like status of the classical concepts.

Potentially problematic for my reading, however, are statements like the
following: ``... no experience is definable without a logical frame
and ... any apparent disharmony can be removed only by an appropriate
widening of the conceptual framework.'' \citep[p. 82]{bohr1954}, which
lead Kaiser to write, of Bohr's view:

\begin{quote}
... there is also a very un-Kantian sentiment expressed in the end of
Bohr's quotation: our  formal frame might need to be
\emph{altered}. ... Kant viewed this formal frame, which includes the
forms of intuition and the categories, as \emph{a priori} and
unalterable. Bohr followed a two-faculty format but he rejected
\emph{a priorism}. \citep[pp. 222-223]{kaiser1992}
\end{quote}

Yet Kaiser's interpretation is misleading, at best, for it seems to
conflate Bohr's view with Heisenberg's. Heisenberg maintained that the
gradual evolution of scientific concepts (or even the human species)
would allow us to transcend our limitation to the classical concepts
(\emph{Cf.} \citealt[p. 83]{heisenberg1959},
\citealt[p. 124]{heisenberg1971}). This was not Bohr's view: ``... it
would be a misconception to believe that the difficulties of the
atomic theory may be evaded by eventually replacing the concepts of
classical physics by new conceptual forms.''
\citeyearpar[p. 16]{bohr1934}. And again:

\begin{quote}
We must, in fact, realise that the unambiguous interpretation of
any measurement must be essentially framed in terms of the classical
physical theories, and we may say that in this sense the language of
Newton and Maxwell will remain the language of physicists for all
time. \citeyearpar[p. 692]{bohr1931}.
\end{quote}

What Bohr means by `widening', then, is not a fundamental alteration of
our basic conceptual framework, but an imaginative use of our
framework's own resources in order to extend its reach. ``Indeed, the
development of atomic physics has taught us how, \emph{without leaving
  common language}, it is possible to create a framework sufficiently
wide for an exhaustive description of new experience.''
\citeyearpar[p. 88, emphasis mine]{bohr1955}.

Specifically related to Kant, Folse objects that Kant was a
`subjectivist' philosopher, while Bohr's intention was to provide an
objective description of experience. Folse writes:

\begin{quote}
These facts have given rise to the view held by some of the most
perceptive of Bohr's interpreters that his position contains Kantian
elements supporting a subjectivistic reading of complementarity. Since
Bohr specifically stated complementarity provides an objective
description of experience, it would seem that virtually any such
reading would be contrary to his intent
... \citep[p. 217]{folse1985}.
\end{quote}

But this misinterprets Kant. Kant's theoretical philosophy, as we have
seen, revolves around the question of how to give an objective
description of experience; thus he takes great pains,
for instance, to distinguish the \emph{objective} succession of
appearances from the \emph{subjective} one. If not ascribing to
na\"ive realism amounts to being a subjectivist then Kant is
guilty on all counts, however I do not think this is the type of
subjectivism that Folse is referring to, for Bohr would be guilty of
this charge as well. For Kant, possible experience is constrained by
the forms of our intuition, space and time, and by the concepts by
which we are able to combine these intuitions into one representation
of an object. But this is no different from Bohr's insistence that we
require classical concepts for the unambiguous description of
experience.

Bohr was known to have admired the work of the American pragmatist
William James, and this has been taken by Folse
(\citealt[p. 49-51]{folse1985}, \citealt[pp. 217-221]{folse1985}) to
tell against a Kantian influence on Bohr, for James was sharply
critical of Kant. As Kaiser points out, however, James' criticisms of
Kant are all directed at Kant's a priorism and not at the other
aspects of his philosophy. This is perfectly compatible with a
picture of Bohr as accepting certain aspects of Kant's philosophy
while rejecting others. It is certainly not without precedent for one
philosopher to be influenced by two rivals: Kant himself was strongly
influenced by both Newton and Leibniz; their rivalry did not stop him
from incorporating aspects of both of their views into his own.

Indeed, many philosophers have borrowed from Kant without making
themselves into carbon copies. The Neo-Kantian philosopher, Ernst
Cassirer, for instance, rejects the a priori status of Kant's
classical concepts \citeyearpar[pp. 194-195]{cassirer1936} while still
maintaining a broadly Kantian epistemology; the intuitionist
mathematician L.E.J. Brouwer was strongly influenced by
Kant\textemdash Brouwer, like Kant, founds arithmetic on the pure
intuition of time\textemdash yet Brouwer rejects the pure intuition of
space in light of the development of non-Euclidean geometry. Both
Reichenbach and Carnap began their careers as Neo-Kantians before
turning towards logical empiricism in light of the developments in
geometry and logic \citep[]{friedman2000,
glymour-eberhardt2008}. Frege, throughout his career, though
critical of Kant's views on arithmetic, nevertheless believed Kant to
be correct for the case of geometry, even in the wake of the modern
developments.\footnote{\emph{Cf.} \citet*[]{merrick2006} for more on the
  relation between Kant and Frege.} After mercilessly skewering most
of his own contemporaries and predecessors, Frege writes, of Kant: ``I
have no wish to incur the reproach of picking petty quarrels with a
genius to whom we must all look up with grateful awe; I feel bound,
therefore, to call attention also to the extent of my agreement with
him, which far exceeds any disagreement.'' \citep[\textsection
89]{frege1980}. Brouwer's arch-rival, Hilbert, was also influenced
by Kant. Hilbert, in the epigraph to his \emph{Foundations of
Geometry}, quotes Kant: ``All human knowledge begins with
intuitions, thence passes to concepts and ends with ideas.''
\citep{hilbert1902}. All of these thinkers incorporated parts of
Kantian philosophy into their own. Bohr was a contemporary of all of
these men; further, he had access to Kantian ideas through his
lifelong friend and mentor, Harald H\o ffding, who was something of a
Kant scholar. Consider H\o ffding's analysis of Kantian philosophy, in
light of our discussion of Complementarity:

\begin{quote}
Experience not only implies that we conceive something in space and
time, but likewise that we are able to combine what is given in space
and time in a definite way, i.e. as indicated in the concepts of
magnitude and causality. This is the only means of distinguishing
between experience and mere representation or imagination. All
extensive and intensive changes must proceed continuously,
i.e. through every possible degree of extension and intensity,
otherwise we could never be certain of having any real
experience. Gaps and breaks must be impossible (non datur hiatus non
datur saltus). The origin of each particular phenomenon moreover must
be conditioned by certain other phenomena, ... Wherever there appear
to be gaps in the series of perceptions we assume that further
investigation will discover the intervening members. This
demonstration of the \emph{validity of the categories} of magnitude
and causality likewise involves a limitation: The validity of the
categories can only be affirmed within the range of possible
experience; they cannot be applied to things which from their very
nature cannot become objects of
experience. \citep[147-148]{hoffding1922}.
\end{quote}

A last objection that I will address, before concluding, is with
regards to the common misconception of Bohr as a positivist. This
conception of Bohr has been popularised by, among others, Karl Popper
and Mario Bunge. I will not spend much time answering it here. In
addition to directing the interested reader to Don Howard's
illuminating article \citeyearpar{howard2004} on the subject, I
will simply point out that this is a view that Bohr (as quoted by
Heisenberg) explicitly denied: ``Positivist insistence on conceptual
clarity is, of course, something I fully endorse, but their
prohibition of any discussion of the wider issues, simply because we
lack clear-cut enough concepts in this realm, does not seem very
useful to me\textemdash this same ban would prevent our understanding
of quantum theory.'' \citep[p. 208]{heisenberg1971}.

One may, of course, ignore Bohr's own words here and presume to
understand him better than he understood himself. If one were to make
such a claim, it would not be objectionable as such; however, given the
current, and widely acknowledged, dearth of understanding with respect
to Bohr's views on quantum mechanics, such a presumption should be
regarded as highly dubious.

In this paper I have highlighted the parallels between Bohr's
doctrine of Complementarity and Kant's theoretical philosophy. We
have seen how Bohr's principle of complementarity and Kant's
theoretical philosophy are common in their approach: that both
approaches are centred around what each thinker took to be the limits
of objective experience. We have seen how, in order to transcend these
limits, Bohr appealed to what a Kantian would call noumena in the
positive sense, or ideas of reason. We have seen how a Kantian (who
does not deny the validity of the uncertainty relations), starting
from the principles of Kantian philosophy, would be led to many of the
same conclusions as Bohr. Finally, we have seen how the objections to
the link between the two thinkers rest on either a misinterpretation
of Kant, or on a misrepresentation of Bohr, or both.

Complementarity is the natural outcome of a broadly Kantian
epistemological framework and a Kantian approach to natural science,
conjoined with Heisenberg's uncertainty relations. There is a very
strong similarity in spirit, if not in technical detail, between
Bohr's and Kant's approaches to natural science, and I hope to have
inspired the conviction that the further examination of these
similarities (and differences) will lead us to a better understanding
of both of these men.

\bibliographystyle{apa-good}
\bibliography{Bibliography}{}

\end{document}